\documentclass[10pt,letterpaper]{article}
\usepackage{opex3}
\usepackage{cite}
\usepackage{graphicx}
\usepackage{amsmath,amsfonts,amssymb}

\setkeys{Gin}{width=\linewidth}

\begin{document}

\title{The selection rule of graphene in a composite magnetic field}

\author{Y. C. Ou,$^1$ Y. H. Chiu,$^{2,4}$ P. H. Yang,$^{3,5}$ and M. F. Lin$^{1,^\ast}$}

\address{
$^1$Department of Physics, National Cheng Kung University, Tainan, Taiwan \\
$^2$National Center for Theoretical Sciences, National Tsing Hua University, Hsinchu, Taiwan\\
$^3$National Center for High-Performance Computing (South), Tainan, Taiwan\\
}

\email{$^4$airegg.py90g@nctu.edu.tw}
\email{$^5$yangp@nchc.narl.org.tw}
\email{$^\ast$mflin@mail.ncku.edu.tw}




\begin{abstract}
The generalized tight-binding model with exact diagonalization method is
developed to calculate the optical properties of monolayer graphene in the
presence of composite magnetic fields. The ratio of the uniform magnetic
field and the modulated one accounts for a strong influence on the
structure, number, intensity and frequency of absorption peaks, and thus the
extra selection rules that are subsequently induced can be explained. When
the modulated field increases, each symmetric peak, under a uniform magnetic
field, splits into a pair of asymmetric peaks with lower intensities. The
threshold absorption frequency exhibits an obvious evolution in terms of a
redshift. These absorption peaks obey the same selection rule that is
followed by Landau level transitions. Moreover, at a sufficiently strong
modulation strength, the extra peaks in the absorption spectrum might arise
from different selection rules.
\end{abstract}











\section{Introduction}

Monolayer graphene (MG), constructed from a single layer of carbon atoms
densely packed in a hexagonal lattice, was successfully produced by
mechanical exfoliation [1, 2] and chemical vapor deposition [3-7]. This
particular material constitutes an excellent system for studying
two-dimensional (2D) physical properties, e.g., quantum Hall effects [8-11].
In the low-energy region $|E^{c,v}|\leq 1$ eV, MG possesses isotropic linear
bands crossing at the $\mathbf{K}$ ($\mathbf{K}^{\prime }$) point and is
regarded as a 2D zero-gap semiconductor, where $c$ ($v$) indicates the
conduction (valence) bands [12]. The linear bands are symmetric about the
Fermi level ($E_{F}=0$) and become nonlinear and anisotropic at higher
energy $|E^{c,v}|>1$ [12]. Most importantly, the quasiparticles related to
the linear bands can be described by the Dirac-like Hamiltonian [13] that is
associated with relativistic particles and dominates the low-energy physical
properties [11, 14, 15]. This special electronic structure has been verified
by experimental measurements [2, 16].

MG has become a potential material candidate for nano-devices due to its
exotic electronic properties. A good understanding of the behavior of MG
under external fields [17-27] is useful for improving the characteristics of
graphene-based nano-devices. In the presence of a uniform perpendicular
magnetic field, the linear bands change into dispersionless Landau levels
(LLs) which obey the specific relationship $E^{c,v}\propto \sqrt{n^{c,v}B_{0}%
}$, where $n^{c}$ ($n^{v}$) is the quantum number of the conduction
(valence) states and $B_{0}$ is the magnetic field strength. The related
anomalous quantum Hall effects [11, 28, 29] and particular optical
excitations [17] have been verified experimentally [10, 11]. The anisotropic
behavior of dispersive quasi-Landau levels (QLLs) and the special selection
rules presented in the related optical absorption spectra are shown for a
modulated magnetic field. Furthermore, Haldane predicted the existence of
quantum Hall effects in MG when it is subjected to the modulated magnetic
field even without any net magnetic flux through the entire space [9]. For
two cases of composite fields, a uniform magnetic field combined with a
modulated magnetic field and a uniform magnetic field combined with a
modulated electric potential, the LL properties are drastically changed by
the modulated field [30-34]. The broken symmetry, displacement of the
localization center, and alteration of the amplitude of the LL wave
functions can, in some cases, be observed [35, 36].

MG is predicted to exhibit feature-rich optical absorption spectra. The
spectral intensity is proportional to the frequency, but no prominent peak
exists at low frequency [37]. However, a uniform perpendicular magnetic
field can lead to a number of symmetric absorption peaks originating from
LLs. Each peak obeys the specific selection rule $\Delta n=|n^{c}-n^{v}|=1$,
which has been confirmed by magneto-transmission measurements [16, 17, 38].
The selection rule is ascribed to the spatial symmetric configuration of the
magneto-electronic wave function. Under a modulated magnetic field, the
optical spectra exhibit many asymmetric principal peaks and subpeaks. The
former satisfy a selection rule similar to those pertinent to LLs, while the
latter do not [39, 40]. For a better understanding, it is necessary to
investigate the primary features of optical excitations in composite
magnetic fields with various field strength ratios between the uniform and
the modulated fields.

The generalized tight-binding model with exact diagonalization method is
developed to resolve the optical absorption spectra. By rearranging the
tight-binding functions, the giant Hermitian matrix corresponding to the
experimental fields can be transformed into a band-like matrix, so that the
numerical computation time is greatly reduced [41-44]. Moreover, the $\pi $%
-electronic structure of MG can be solved in the wide energy range of 5 eV,
a solution proven valid regardless of whether a magnetic, electric or
composite field is applied. In this work, we focus on the optical absorption
spectra of MG under a composite magnetic field. By means of controlling the
ratio between a uniform magnetic field and a modulated one, the
magneto-optical properties can be thoroughly explored and then presented in
detail. For an increased modulated field, each symmetric peak, under a
uniform magnetic field, splits into a pair of asymmetric peaks. Redshift is
an obvious evolution exhibited by the threshold absorption frequency.
Perceivably, each peak obeys the same selection rule as in the situation
where only the uniform magnetic field exists until the modulated field is
increased beyond a certain point. As a result, important differences of the
absorption peaks are observed with regard to their structure, number,
intensity, frequency and selection rule when the applied magnetic field is
varied among the uniform, composite and purely modulated ones.

\section{Band-like Hamiltonian matrix in external fields}

\begin{figure}[tbp]
\begin{center}
\includegraphics[]{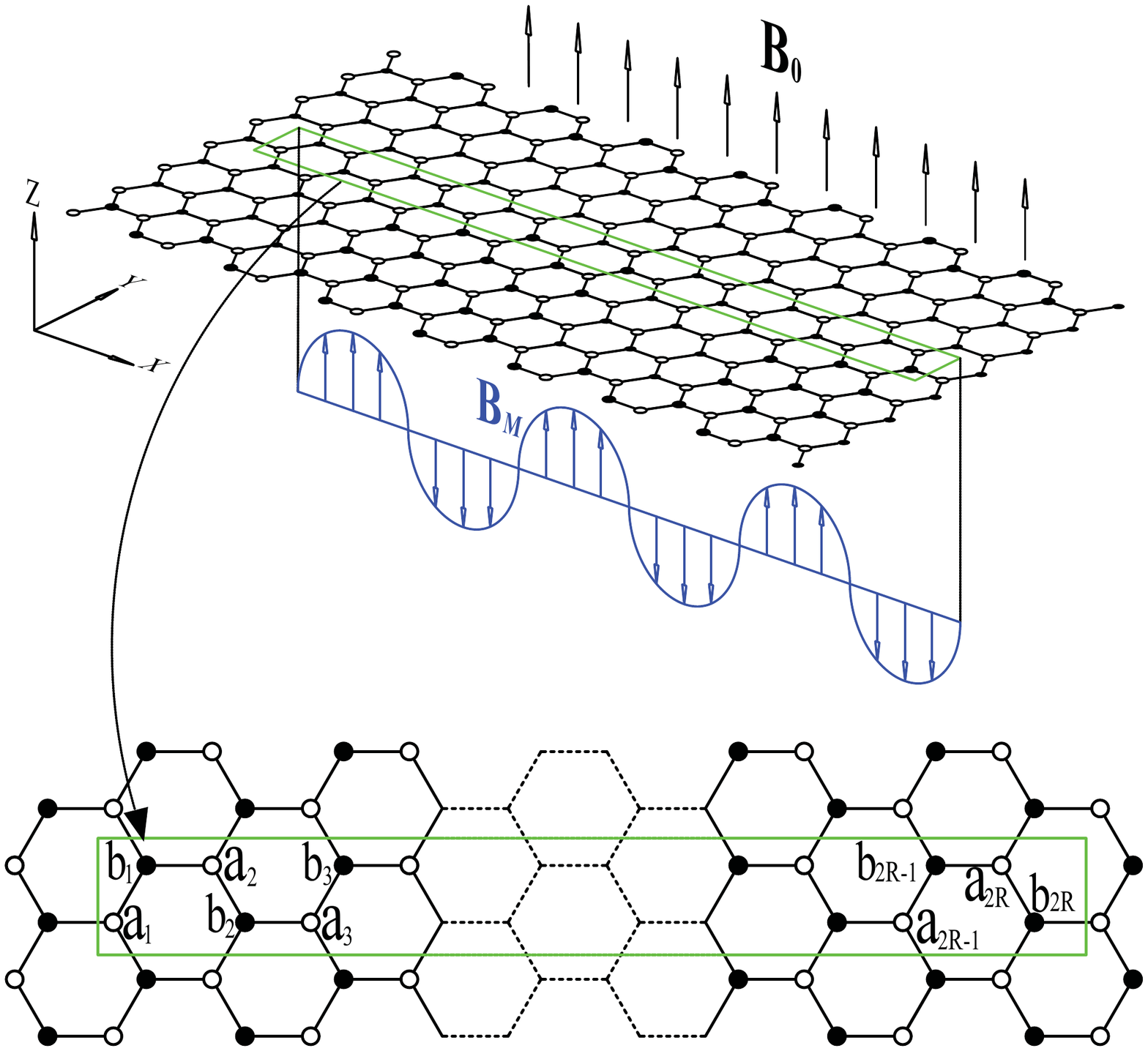}\\[0pt]
\end{center}
\caption{ The primitive cell of a monolayer graphene in a uniform magnetic
field and a spatially modulated magnetic field along the armchair direction.}
\label{fig1.eps}
\end{figure}

The low-frequency optical properties of MG are determined by the $\pi $%
-electronic structure resulting from the 2$p_{z}$ orbitals of the carbon
atoms. The generalized tight-binding model with the exact diagonalization
method is developed to characterize the electronic properties, and then the
gradient approximation is applied to obtain the optical-absorption spectra.
In the presence of magnetic fields, the vector potential induces Peierls
phases [39, 41, 45, 46], which are then accumulated in the Bloch wave
functions. Such a phase owns the extra period $R$ or changes the unit cell.
The enlarged rectangular unit cell, marked by the green rectangle in Fig. 1,
is chosen as the primitive unit cell. In this work, the major discussions
focus on $R$ along the armchair direction. MG is subjected to three kinds of
magnetic fields: a uniform perpendicular magnetic field, a periodically
modulated magnetic field, and a composite magnetic field comprised of both
the uniform and modulated parts. In the case of the uniform magnetic field $%
\mathbf{B}_{0}=B_{0}\widehat{z}$, the period is given by $R=R_{0}=\phi
_{0}/(3\sqrt{3}b^{\prime 2}B_{0}/2)$, where $b^{\prime }$=1.42 \r{A}\ is the
C-C bond length, and $\phi _{0}$ ($=hc/e=4.1356\times 10^{-15}$ [T/m$^{2}$])
is the unit flux quantum. In the case of the modulated magnetic field, $%
\mathbf{B}_{M}=B_{M}\sin (2\pi x/l_{M})$\ $\widehat{z}$ is exerted along the
armchair direction, where $B_{M}$ is the field strength and $l_{M}$ is the
period length with the modulation period $R=R_{M}=l_{M}/3b^{\prime }$. In
the case of the composite field, the period $R=R_{C}$ is the least common
multiple of $R_{0}$\ and $R_{M}$. An enlarged rectangular unit cell induced
by an external field encompasses $2R$ $a$ atoms and $2R$ $b$ atoms. The
Hamiltonian matrix is a $4R\times 4R$ Hermitian\ matrix spanned by $4R$ TB
functions. Based on the arrangement of odd and even atoms in the primitive
cell, the Bloch wave function $|\Psi _{\mathbf{k}}\rangle $\ can be
expressed as:
\begin{equation}
|\Psi _{\mathbf{k}}\rangle =\sum\limits_{m=1}^{2R-1}(A_{\mathbf{o}%
}^{c,v}|a_{m\mathbf{k}}\rangle +B_{\mathbf{o}}^{c,v}|b_{m\mathbf{k}}\rangle
)+\sum\limits_{m=2}^{2R}(A_{\mathbf{e}}^{c,v}|a_{m\mathbf{k}}\rangle +B_{%
\mathbf{e}}^{c,v}|b_{m\mathbf{k}}\rangle )\text{,}
\end{equation}%
where $|a_{m\mathbf{k}}\rangle $ ($|b_{m\mathbf{k}}\rangle $) is the TB
function corresponding to\ the 2$p_{z}$ orbital of the $m$th $a$ ($b$) atom.
$A_{\mathbf{o}}^{c,v}$ ($A_{\mathbf{e}}^{c,v}$) and $B_{\mathbf{o}}^{c,v}$ ($%
B_{\mathbf{e}}^{c,v}$) are the subenvelope functions representing the
amplitudes of the wave functions of the $a$- and $b$-atoms respectively,
where $o$ ($e$) represents an odd (even) integer. Since the features of $%
A_{o}^{c,v}$ ($B_{o}^{c,v}$) and $A_{e}^{c,v}$ ($B_{e}^{c,v}$)\ are similar,
choosing only the amplitudes $A_{\mathbf{o}}^{c,v}$ and $B_{\mathbf{o}}^{c,v}
$ suffices to understand the electronic and optical properties. The $%
4R\times 4R$ Hamiltonian matrix, which determines the magneto-electronic
properties, is a giant Hermitian matrix with respect to the external fields
actually used in the experiments. To make the calculations more efficient,
the matrix is transformed into an $M\times 4R$ band-like matrix by a
suitable rearrangement of the tight-binding functions, where $M$ is much
smaller than $4R$. For example, one can arrange the basis functions
according to the following sequence $|a_{1\mathbf{k}}\rangle $, $|b_{2R%
\mathbf{k}}\rangle $, $|b_{1\mathbf{k}}\rangle $, $|a_{2R\mathbf{k}}\rangle $%
, $|a_{2\mathbf{k}}\rangle $, $|b_{2R-1\mathbf{k}}\rangle $, $|b_{2\mathbf{k}%
}\rangle $, $|a_{2R-1\mathbf{k}}\rangle $, ......$|a_{R-1\mathbf{k}}\rangle $%
, $|b_{R+2\mathbf{k}}\rangle $, $|b_{R-1\mathbf{k}}\rangle $, $|a_{R+2%
\mathbf{k}}\rangle $, $|a_{R\mathbf{k}}\rangle $, $|b_{R+1\mathbf{k}}\rangle
$, $|b_{R\mathbf{k}}\rangle $; $|a_{R+1\mathbf{k}}\rangle $. By performing
these calculations, the nonzero Hamilatonian\ matrix\ elements can be
formulated as%
\begin{equation}
\langle b_{m^{\prime }\mathbf{k}}|H|a_{m\mathbf{k}}\rangle
=[t_{1k}(m)+t_{2k}(m)]\delta _{m^{\prime },m}+t_{3k}(m)\delta _{m^{\prime
},m-1}\text{,}
\end{equation}%
where $t_{1k}(m)=\gamma _{0}\exp [i(k_{x}b^{\prime }/2+k_{y}\sqrt{3}%
b^{\prime }/2-G_{0}-G_{M})]$, $t_{2k}(m)=\gamma _{0}\exp [i(k_{x}b^{\prime
}/2-k_{y}\sqrt{3}b^{\prime }/2-G_{0}-G_{M})]$ and $t_{3k}(m)=\gamma _{0}\exp
[-i(k_{x}b^{\prime })]$ are the three nearest-neighbor hopping integrals. $%
\gamma _{0}=2.5$ eV is the nearest-neighbor interaction. $G_{0}=-\frac{\pi
\lbrack (m-1)+1/6]}{R_{0}}$ and $G_{M}=\frac{6(R_{M})^{2}}{\pi }\frac{\phi
^{\prime }}{\phi _{0}}\cos [\frac{\pi }{R_{M}}(m-\frac{5}{6})]\sin (\frac{%
\pi }{6R_{M}})$ are the Peierls phases in the off-diagonal elements
associated with the uniform and modulated magnetic fields, respectively.

When electrons are excited from occupied valence to unoccupied conduction
bands by an electro-magnetic field,\ only inter-$\pi $-band excitations
exist in the zero temperature case. Based on Fermi's golden rule, the
optical absorption function results in the following form
\begin{eqnarray}
A(\omega ) &\propto &\sum\limits_{c,v,\widetilde{n},\widetilde{n}^{\prime
}}\int_{1stBZ}\frac{d\mathbf{k}}{(2\pi )^{2}}\left\vert \langle \Psi ^{c}(%
\mathbf{k},n)|\frac{\widehat{\mathbf{E}}\cdot \mathbf{P}}{m_{e}}|\Psi ^{v}(%
\mathbf{k},n^{\prime })\rangle \right\vert ^{2}  \notag \\
&&\times \mathrm{Im}\left[ \frac{f(E^{c}(\mathbf{k},n))-f(E^{v}(\mathbf{k}%
,n^{\prime }))}{E^{c}(\mathbf{k},n)-E^{v}(\mathbf{k},n^{\prime })-\omega
-i\Gamma }\right] \text{,}
\end{eqnarray}%
where $f(E(\mathbf{k},\widetilde{n}))$ is the Fermi-Dirac distribution
function, $\Gamma $ ($=2\times \,10^{-4}\gamma _{0}$) is the broadening
parameter, $\mathbf{P}$ is the momentum, and $m_{e}$ is the bare electron
mass. Meanwhile, $n$ and $n^{\prime }$ are the quantum numbers with respect
to the initial and final states. The electric polarization $\widehat{\mathbf{%
E}}$ along $\widehat{y}$ is taken into account for discussions. Within the
gradient approximation [47-51], the velocity matrix element $M^{cv}=\langle
\Psi ^{c}(\mathbf{k},n)|\frac{\widehat{\mathbf{E}}\cdot \mathbf{P}}{m_{e}}%
|\Psi ^{v}(\mathbf{k},n^{\prime })\rangle $ is formulated as%
\begin{equation}
\sum\limits_{m,m\prime =1}^{2R_{C}}[(A_{\mathbf{o}}^{c}{}+A_{\mathbf{e}%
}^{c})^{\ast }\times (B_{\mathbf{o}}^{v}+B_{\mathbf{e}}^{v})]\nabla _{%
\mathbf{k}}\langle a_{m\mathbf{k}}|H|b_{m^{\prime }\mathbf{k}}\rangle +h.c.%
\text{.}
\end{equation}%
Equation (4) implies that the main features of the wave functions are key
factors in determining the selection rules and the absorption intensity of
the optical excitations. Similar gradient approximations have been
successfully applied to investigate optical spectra of carbon-related
systems, e.g., graphite [47], graphite intercalation compounds [51], carbon
nanotubes [52], few-layer graphenes [53], and graphene nanoribbons [54].

\section{Uniform magnetic field combined with a modulated magnetic field}

\begin{figure}[tbp]
\begin{center}
\includegraphics[]{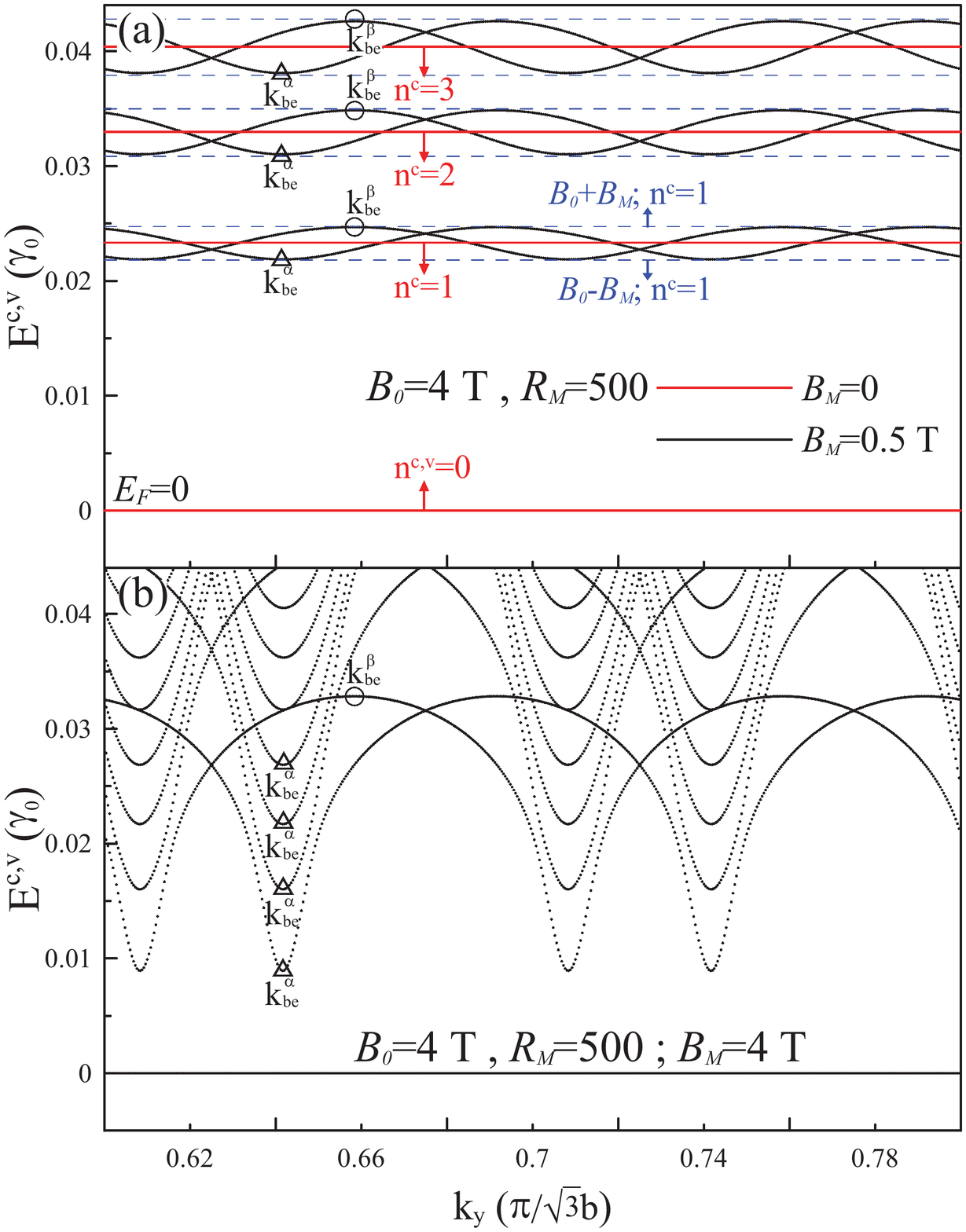}\\[0pt]
\end{center}
\caption{ The low $k_{y}$-dependent energy bands for $B_{M}\leq B_{0}$ case\
under (a) the uniform magnetic field $B_{0}=4$ T (red curves) and the
composite field $B_{0}=4$ T in conjunction with $R_{M}=500$ and $B_{M}=0.5$
T (black curves), and (b) $B_{0}=4$ T in conjunction with $R_{M}=500$ and $%
B_{M}=4$ T. The triangular and circular symbols correspond to the band-edge
states $k_{be}^{\protect\alpha }$ and $k_{be}^{\protect\beta }$,
respectively.}
\label{fig2.eps}
\end{figure}

Low-energy band structures of MG exhibit rich features in the presence of
magnetic fields. The uniform perpendicular magnetic field causes the states
to congregate and induces dispersionless Landau levels (LLs), indicated by
the red curves in Fig. 2 for\ $B_{0}=4$ T. The unoccupied LLs and occupied
LLs\ are symmetric about the Fermi level ($E_{F}=0$). Each LL is
characterized by the quantum number\ $n^{c,v}$, which corresponds to the\
number of zeros in the eigenvectors of harmonic oscillators [42, 55]. Each
LL is fourfold degenerate for each ($k_{x}$, $k_{y}$) state without
considering the spin degeneracy. Its energy can be approximated by a simple
square-root relationship $|E_{n}^{c,v}|\propto \sqrt{n^{c,v}B_{0}}$ [8, 56],
which is valid only for $|E_{n}^{c,v}|\leq 1$ eV [8].

The main characteristics of the LLs are affected by the modulated magnetic
field, as shown in Fig. 2(a) by the black curves for $B_{M}=0.5$ T and $%
R_{M}=500$. Only the $n^{c,v}=0$ LL remains unchanged, i.e., it retains its
fourfold degeneracy at $E_{F}=0$. On the other hand, each dispersionless LL
with $n^{c,v}\geqslant 1$ splits into two periodic oscillation subbands with
double degeneracy. The subbands possess two kinds of band-edge states, $%
k_{be}^{\alpha }$ and $k_{be}^{\beta }$, which are associated with the
minimum field strength $B_{0}-B_{M}$ and the maximum field strength $%
B_{0}+B_{M}$, respectively. At a small modulation strength ($B_{M}/B_{0}\leq
1/8$), the subbands oscillate between two LLs at the field strengths $%
B_{0}\pm B_{M}$, as shown by the blue dashed lines in Fig. 2(a). The
minimum and maximum of the oscillation subbands are proportional to $\sqrt{%
B_{0}-B_{M}}$ and $\sqrt{B_{0}+B_{M}}$, respectively. The surrounding
electronic states at $k_{be}^{\beta }$ congregate more easily, which results
in a smaller band curvature with a larger density of states (DOS). On the
contrary, fewer states congregate at $k_{be}^{\alpha }$, a situation leading
to the larger band curvature and a smaller DOS.

The band structure may be considerably modified by increasing the strength
of the modulated magnetic field. As $B_{M}$ increases to the magnitude of $%
B_{0}$, more complex energy spectra are\ introduced, as shown in Fig. 2(b)
for $R_{M}=500$ and $B_{M}=4$ T. The oscillation subbands with $%
n^{c,v}\geqslant 1$ display wider oscillation amplitudes, stronger energy
dispersions, and greater band curvatures. The strong oscillatory subbands
with different quantum numbers overlap with one another, and the subband
amplitudes are almost linearly magnified by $B_{M}$. The wave vectors
associated with the band-edge states have no dependence on $B_{M}$.\ The
greatest and smallest band curvatures occur at the local minimum $%
k_{be}^{\alpha }$ and the local maximum $k_{be}^{\beta }$, respectively.
This implies that there are more $k_{be}^{\alpha }$ and fewer $k_{be}^{\beta
}$ in the lower state energies. However, a simple relation between the
subband amplitudes and $\sqrt{B_{0}\pm B_{M}}$ is absent.

\begin{figure}[tbp]
\begin{center}
\includegraphics[]{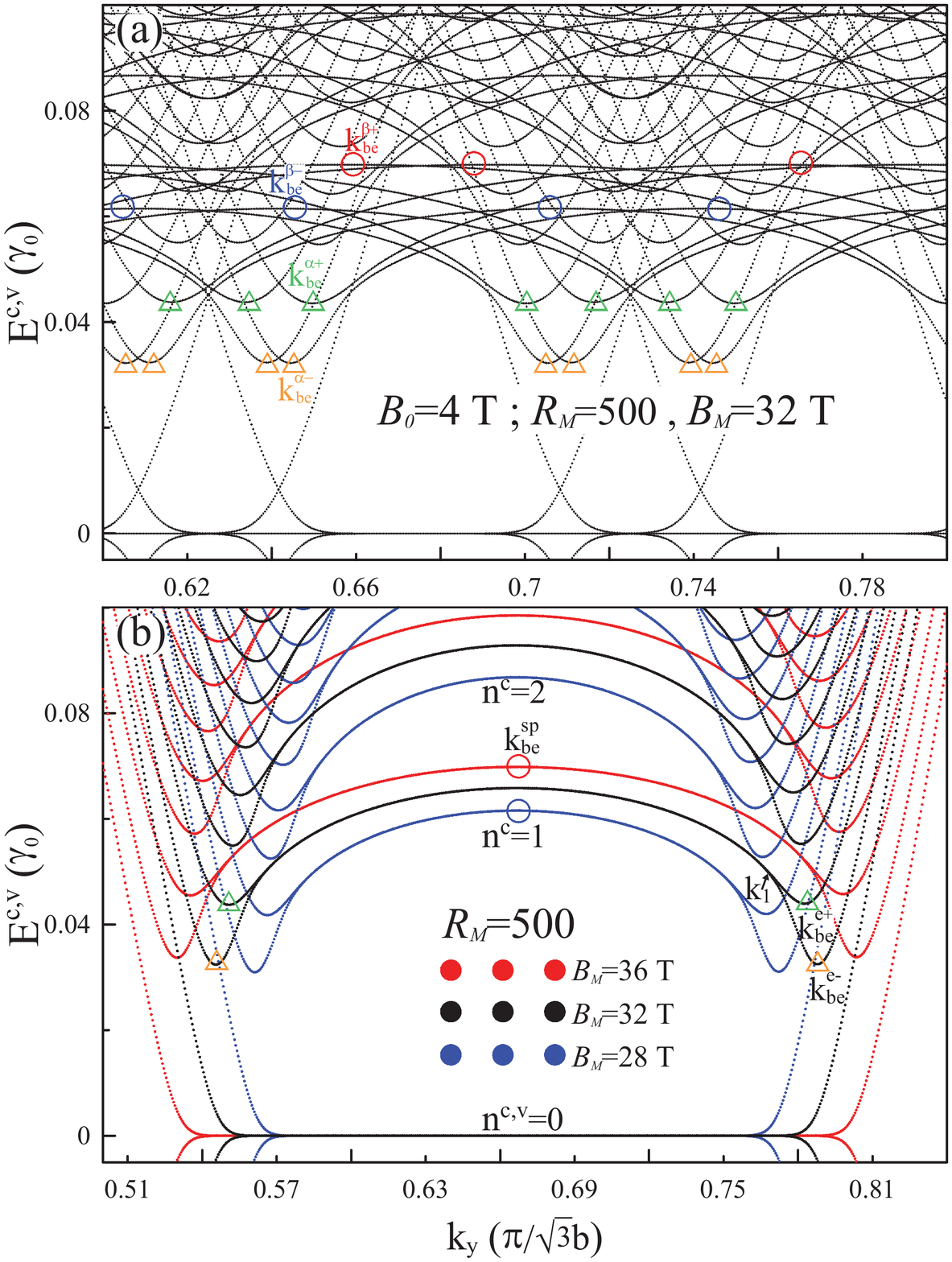}\\[0pt]
\end{center}
\caption{ The low $k_{y}$-dependent energy bands for the $B_{M}>B_{0}$ case\
at (a) the composite field $B_{0}=4$ T in conjunction with $R_{M}=500$ and $%
B_{M}=32$ T, and (b) the pure modulated magnetic field $R_{M}=500$, $B_{M}=$%
36, 32 and 28 T (red, black and blue curves, respectively). The triangular
and circular symbols represent the same meanings as in Fig.2.}
\label{fig3.eps}
\end{figure}

An increase in the strength of the modulated field, as $B_{M}\gg B_{0}$, can
drastically change the band structure. Figure 3(a) illustrates the complex
oscillatory subbands for $R_{M}=500$ and $B_{M}=32$ T . The oscillation
amplitudes do not demonstrate a simple relationship with $B_{M}$. It should
be noted that neither the minima of the conduction bands nor the maxima of
the valence bands exceed $E_{F}=0$. Thus, there is no overlap between the
conduction and valence bands, regardless of the modulation strength. The
wave vectors of the band-edge states in the case of a large modulation
strength ($B_{M}/B_{0}\geq 8$) are different from those in the case of $%
B_{M}\leq B_{0}$ (depicted in Fig. 2). Therefore, each type of band-edge
state corresponds to two different energies. The band-edge states belonging
to the $\alpha $ ($\beta $) type own the two energies $k_{be}^{\alpha -}$
and $k_{be}^{\alpha +}$ ($k_{be}^{\beta -}$ and $k_{be}^{\beta +}$), as
indicated by the orange and green triangles (blue and red circles) in the
figure. Furthermore, the band structure in the composite fields displays
band-edge state energies similar to those found under the influence of only
a modulated magnetic field. In the presence of a pure modulated magnetic
field, energy bands are partial flat bands at $E_{F}=0$ and parabolic bands
elsewhere, as shown in Fig. 3(b). Each parabolic subband has one specific
band-edge state $k_{be}^{sp}$ at $k_{y}=2/3$ and four extra band-edge states
$k_{be}^{e+}$'s and $k_{be}^{e-}$'s at both sides of $k_{y}=2/3$ (indicated
by the open circle and triangles), respectively. For the former, the\ energy
of $k_{be}^{sp}$ for the modulation strengths $B_{0}+B_{M}$ and $B_{0}-B_{M}$%
\ is similar to those of $k_{be}^{\beta +}$ and $k_{be}^{\beta -}$,
respectively. The latter $k_{be}^{e+}$ and $k_{be}^{e-}$ for the modulation
strength $B_{M}$ own the same energy as\ $k_{be}^{\alpha +}$ and $%
k_{be}^{\alpha -}$, respectively. The quantum number $n^{c,v}$\
corresponding to the partial flat bands at $E_{F}=0$ is\ defined as zero,
and $n^{c,v}\geq 1$ are associated with the $n$-th ($n=1,2,3\ldots $)
conduction and valence parabolic subbands. The definition of $n^{c,v}$\ is
discussed in the following paragraph and graphically illustrated in Fig. 5.

\begin{figure}[tbp]
\begin{center}
\includegraphics[]{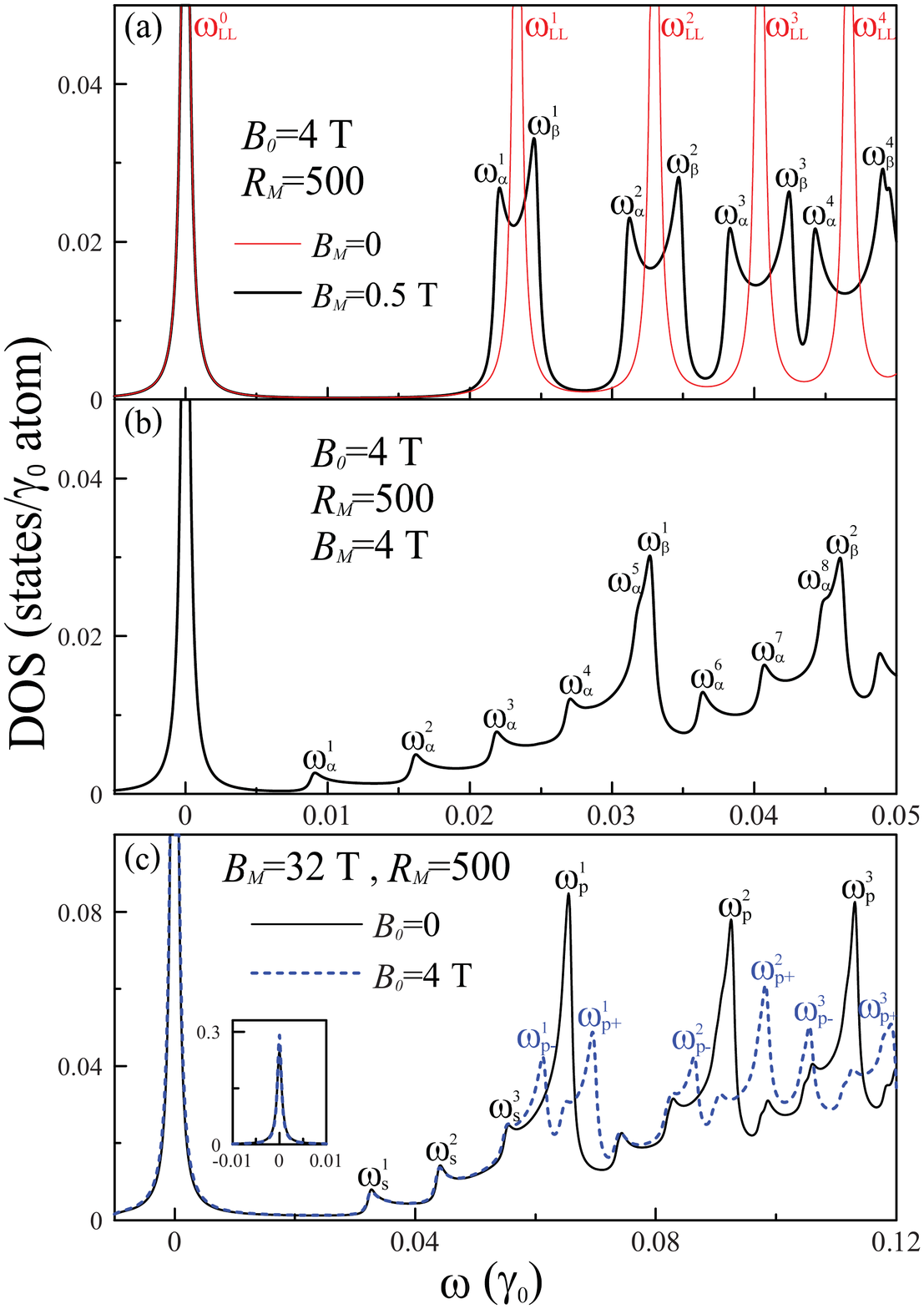}\\[0pt]
\end{center}
\caption{ The low-frequency density of states at (a) the uniform magnetic
field $B_{0}=4$ T (red curves) and the composite field $B_{0}=4$ T in
conjunction with $R_{M}=500$ and $B_{M}=0.5$ T (black curves), (b) $B_{0}=4$
T in conjunction with $R_{M}=500$ and $B_{M}=4$ T, and (c) $B_{0}=4$ T in
conjunction with $R_{M}=500$ and $B_{M}=32$ T (blue curves), and the pure
modulated magnetic field $R_{M}=500$ and $B_{M}=$32 T (black curves).}
\label{fig4.eps}
\end{figure}

The above-mentioned features of band structures influenced by the modulated
magnetic fields would be reflected in the density of states (DOS). In the
presence of a uniform magnetic field, DOS exhibits many delta-function-like
peaks resulting from LLs, as shown by the red curve for $B_{0}=4$ T in Fig.
4(a). Such peaks suggest that LLs possess 0D energy levels. With an applied
modulated magnetic field, each peak, except for the $n^{c,v}=0$ LL at $%
E_{F}=0$, splits into a pair of square-root divergent peaks $\omega _{\alpha
}$ and $\omega _{\beta }$, as shown by the black curve in Fig. 4(a) for $%
R_{M}=500$ and $B_{M}=0.5$ T. The square-root divergence implies that the 0D
energy level becomes the 1D energy band. Each pair of peaks $\omega _{\alpha
}$ and $\omega _{\beta }$ corresponds to the $k_{be}^{\alpha }$ and $%
k_{be}^{\beta }$ states, respectively, and the peak frequencies are
identical to those of LLs at $B_{0}+B_{M}$ and $B_{0}-B_{M}$. However, the
two peaks $\omega _{\alpha }$ and $\omega _{\beta }$ possess different peak
heights owing to the distinct band curvatures at the two kinds of band edge
states. The frequency and intensity are strongly dependent on the modulation
strength. The peak heights decline and the spacing between $\omega _{\alpha
} $ and $\omega _{\beta }$ rises for a large modulation strength $B_{M}=4$
T, as shown in Fig. 4(b). This reflects the fact that the greater
curvatures and wider amplitudes of the oscillation subbands are a result of
the increasing $B_{M}$.

As the modulated field strength is further raised to $B_{M}=32$ T (blue
dashed curves in Fig. 4(c)), the DOS displays some features similar to
those of the DOS in the modulated magnetic field where $R_{M}=500$ and $%
B_{M}=32$ T (black solid curves). The modulated magnetic field alone leads
to a symmetric delta-function-like peak at $\omega =0$ (inset in Fig. 4(c)), as well as asymmetric square-root divergent peaks. The former comes
from the partial flat bands at $E_{F}=0$. The latter can be further divided
into weak subpeaks ($\omega _{s}^{n}$) and strong
principal peaks ($\omega _{p}^{n}$) that are, respectively, dominated by the
extra band-edge states ($k_{be}^{e+}$ and $k_{be}^{e-}$) and specific
band-edge states ($k_{be}^{sp}$). In the cases of a composite magnetic
field, the frequencies and intensities of the subpeaks are almost the same
as those in the modulated magnetic field alone. However, the principal peaks
possess pair structures of $\omega _{P+}^{n}$ and $\omega _{P-}^{n}$, which
respectively correspond to the two different band-edge states $k_{be}^{\beta
+}$ and $k_{be}^{\beta -}$.\

\begin{figure}[tbp]
\begin{center}
\includegraphics[]{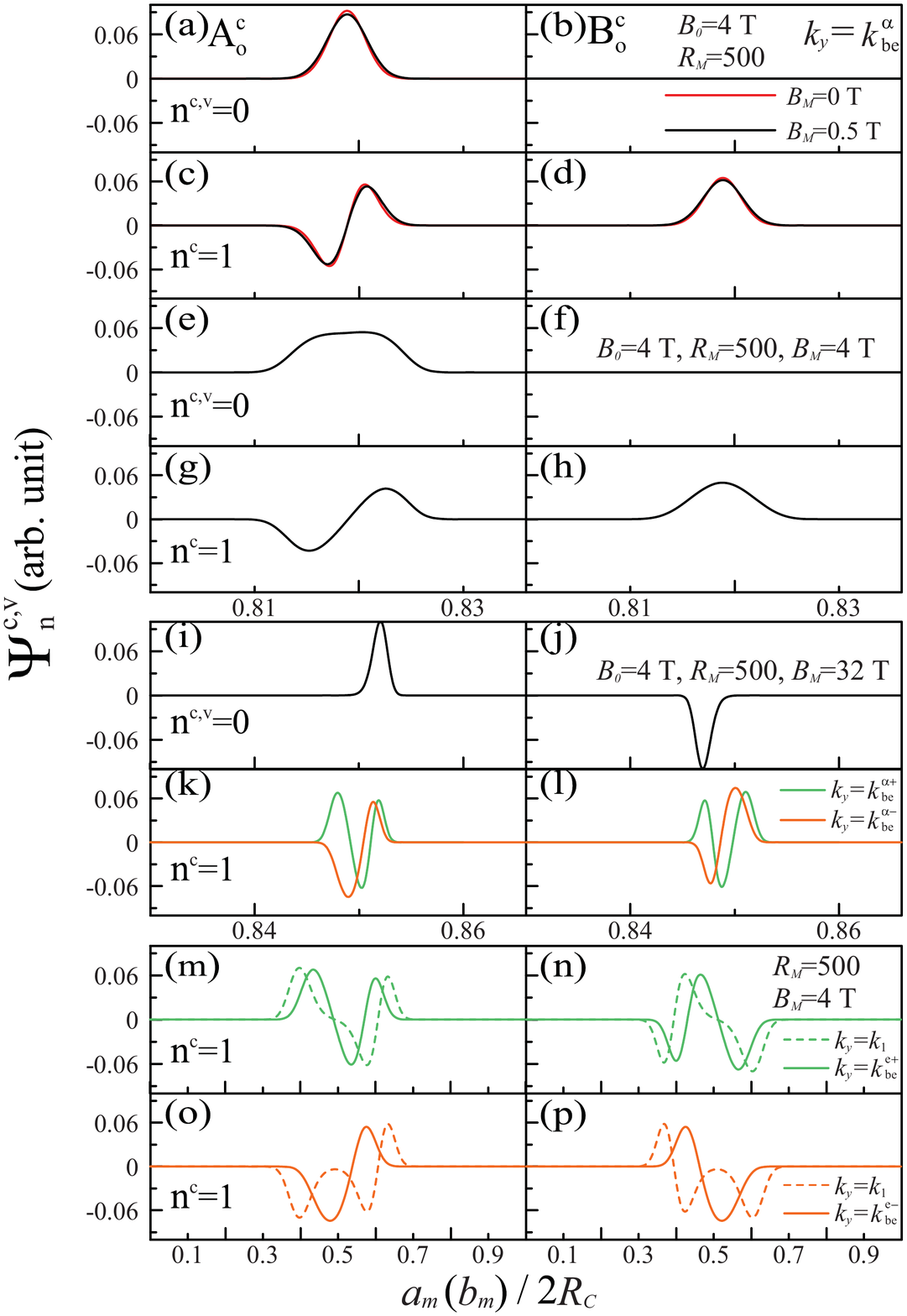}\\[0pt]
\end{center}
\caption{ The wave functions with $n^{c,v}=0$ and $n^{c}=1$ at $k_{be}^{%
\protect\alpha }$ for (a)-(d) the uniform magnetic field $B_{0}=4$ T (red
curves) and the composite field $B_{0}=4$ T together with $R_{M}=500$ and $%
B_{M}=0.5$ T (black curves), (e)-(h) $B_{0}=4$ T combined with $R_{M}=500$
and $B_{M}=4$ T, and (i)-(l) $B_{0}=4$ T combined with $R_{M}=500$ and $%
B_{M}=32$ T. The wave functions with $n^{c}=1$ at $k_{1}$ (solid curves)\
and $k_{be}^{e\pm }$\ (dashed curves)\ in the pure modulated field $%
R_{M}=500 $ and $B_{M}=32$ T, are shown in (m)-(p).}
\label{fig5.eps}
\end{figure}

The LL wave functions exhibit a specific spatial symmetry. Associated with
the odd and even atoms, the wave functions in Eq. 1 have only a phase
difference of $\pi $, that is, $A_{\mathbf{o}}^{c,v}=-A_{\mathbf{e}}^{c,v}$
and $B_{\mathbf{o}}^{c,v}=-B_{\mathbf{e}}^{c,v}$. Therefore, discussing only
the amplitudes $A_{\mathbf{o}}^{c,v}$ and $B_{\mathbf{o}}^{c,v}$ is
sufficient for understanding the main characteristics of the LLs. The LL
wave functions, which is the eigenvector of the simple harmonic oscillation,
can be described by an $n^{c,v}$-th order Hermite polynomial multiplied by a
Gaussian function, as shown in Fig. 5. These wave functions are distributed
around the localization center, that is, at the $5/6$ position of the
enlarged unit cell. Similar localization centers corresponding to the other
degenerate states occur at the $1/6$, $2/6$, and $4/6$ positions. However,
it is adequate to only consider any one center in evaluating the absorption
spectra due to their identical optical responses. A simple relationship
exists between the two sublattices of $a$- and $b$-atoms, i.e., $A_{\mathbf{o%
}}^{c,v}$ of $n^{c,v}$ is linearly proportional to $B_{\mathbf{o}}^{c,v}$ of
$n^{c,v}+1$. Moreover, the conduction and valence wave functions are related
to each other by $A_{\mathbf{o}}^{c}=A_{\mathbf{o}}^{v}$ and $B_{\mathbf{o}%
}^{c}=-B_{\mathbf{o}}^{v}$.

The modulated field has strong effects on the LL wave functions, but plays
no role in the relationship between the valence and conduction states. The
wave functions corresponding to $k_{be}^{\alpha }$, labeled in Fig. 3(a),
are located at the minimum field strength $B_{0}-B_{M}$. They exhibit
slightly broadened and reduced amplitudes, as indicated by the black curves
in Figs. 5(a)-5(d) for a small $B_{M}=0.5$ T. However, the spatial symmetry
of the wave functions remains unchanged. The simple relationship between $A_{%
\mathbf{o}}^{c,v}$ and $B_{\mathbf{o}}^{c,v}$ of the wave functions is
almost preserved. A stronger modulation strength results in greater spatial
changes of the wave functions, as shown in Figs. 5(e)-(f) for $B_{0}=B_{M}=4$
T. An increased broadening and asymmetry of the spatial distributions at $%
n^{c,v}=0$ are revealed. On the other hand, spatial distributions with $%
n^{c,v}\geqslant 1$ are only widened (i.e., $n^{c}=1$ in Figs. 5(g) and
5(h)), whereas their spatial symmetry is retained.

With a large modulation strength $B_{M}\gg $ $B_{0}$, the wave function is
characterized by a narrow distribution and an increased amplitude, as shown
in Figs. 5(i)-5(l). For $n^{c,v}=0$, the symmetry of the wave functions is
almost restored. The amplitude $B_{\mathbf{o}}^{c,v}$, owing to the fact
that it has the same number of zeros as $A_{\mathbf{o}}^{c,v}$, also
contributes to the wave function. This means that carrier transfers occur
between the $a$- and $b$-atoms. For $n^{c,v}\geq 1$, the spatial symmetry is
broken and the zero mode is changed. The wave functions\ at k$_{be}^{\alpha
+}$ and k$_{be}^{\alpha -}$ are different combinations of two subenvelope
functions, of which the location centers are at zero net field strength, as
indicated by the green and orange curves, respectively. These features are
also obtained in the pure modulated magnetic field case. The wave functions
associated with $k_{y}=k_{1}$ for $B_{M}=32$ T\ and $R_{M}=500$\ in Fig.
4(b) have two nearly overlapping subenvelope functions $\phi _{1}$ and $\phi
_{0}$ with respect to 1 and 0 zero points, as shown in Figs. 5(m)-5(p) by
the green and orange dashed curves. The quantum number $n^{c,v}$ is defined
by the larger number of zeros of the subenvelope functions. As the wave
vector moves toward the extra band-edge states $k_{be}^{e+}$ and $%
k_{be}^{e-} $, the two subenvelope functions $\phi _{1}$ and $\phi _{0}$
shift to the center located at zero net field strength and overlap with each
other. For example, the overlapping wave functions $A_{\mathbf{o}}^{c}$
correspond to $k_{be}^{e+}$ and $k_{be}^{e-}$ superpositioned by the
different combinations of $\phi _{0}+\phi _{1}$ and $-\phi _{0}+\phi _{1}$,
indicated by the green and red curves, respectively.

\begin{figure}[tbp]
\begin{center}
\includegraphics[]{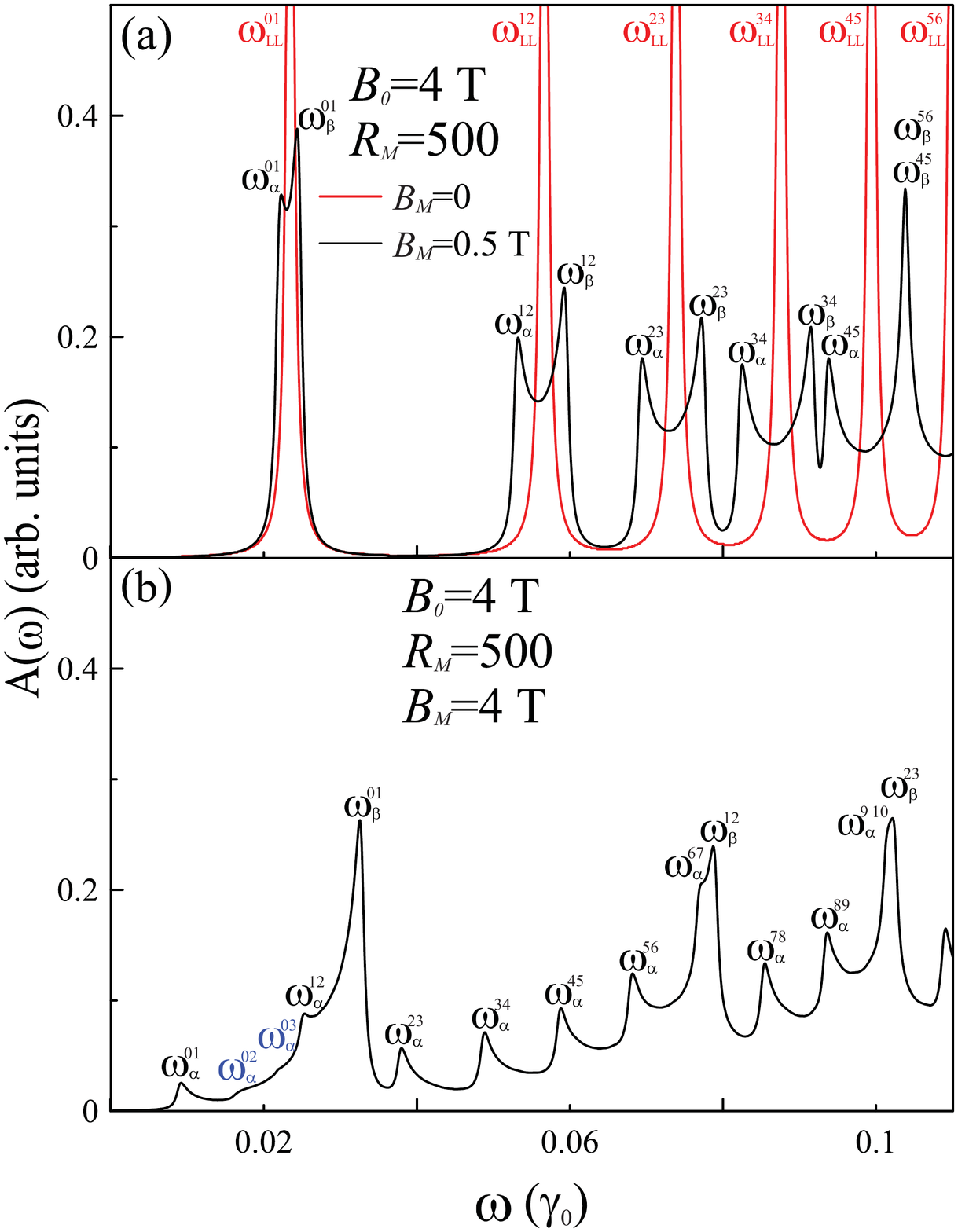}\\[0pt]
\end{center}
\caption{ The low-frequency optical absorption spectra for the $B_{M}\leq
B_{0}$ case corresponding to (a) the uniform magnetic field $B_{0}=4$ T (red
curve) and the composite field $B_{0}=4$ T together with $R_{M}=500$ and $%
B_{M}=0.5$ T (black curve) and (b) the composite field $B_{0}=4$ T combined
with $R_{M}=500$ and $B_{M}=4$ T.}
\label{fig6.eps}
\end{figure}

The low-frequency optical absorption spectrum of the LLs presents numerous
features. Many delta-function-like peaks with an identical intensity exist,
as shown in Fig. 6(a) for $B_{0}=4$ T by the red curves. A single peak $%
\omega _{LL}^{nn^{\prime }}$ comes from two equivalent transition channels:
from the valence LL of $n^{\prime v}$ ($n^{v}$)\ to the conduction LL of $%
n^{c}$ ($n^{^{\prime }c}$). The quantum numbers related to the transition
between the LLs must obey the specific selection rule $\Delta n=|n^{\prime
c(v)}-n^{v(c)}|=1$. The main reason for this is that the velocity matrix $%
M^{cv}$ has a non-zero value only when $A_{\mathbf{o}}^{c,v}$ and $B_{%
\mathbf{o}}^{c,v}$ possess the same number of zeros and it is a dominant
factor\ for the optical excitations of the prominent peaks. However, the
modulated magnetic field modifies the number and intensities of the
absorption peaks. The delta-function-like peak splits into two kinds of
square-root-divergent peaks $\omega _{\alpha }^{nn^{\prime }}$ and $\omega
_{\beta }^{nn^{\prime }}$ with lower intensities, as shown by the black
curves. The divergences of the former and the latter occur near the right-
and left-hand sides of $\omega _{LL}^{nn^{\prime }}$. Each $\omega _{\alpha
}^{nn^{\prime }}$ ($\omega _{\beta }^{nn^{\prime }}$) originates from the
transitions of $k_{be}^{\alpha }(n^{v})\rightarrow k_{be}^{\alpha }(n^{c}+1)$
and $k_{be}^{\alpha }(n^{v}+1)\rightarrow k_{be}^{\alpha }(n^{c})$ [$%
k_{be}^{\beta }(n^{v})\rightarrow k_{be}^{\beta }(n^{c}+1)$ and $%
k_{be}^{\beta }(n^{v}+1)\rightarrow k_{be}^{\beta }(n^{c})$], its absorption
frequency is same as that generated from the LLs at $B_{0}-B_{M}=3.5$ T ($%
B_{0}+B_{M}=4.5$ T). This leads to a redshift for the threshold absorption
frequency. The peak intensity of $\omega _{\alpha }^{nn^{\prime }}$ is lower
than that of $\omega _{\beta }^{nn^{\prime }}$ since the DOS of $%
k_{be}^{\alpha }$ is weaker than that of $k_{be}^{\beta }$. These absorption
peaks obey the selection rule $\Delta n=1$,\ similarly to uniform magnetic
field case. This is due to the fact that the simple relation between $A_{%
\mathbf{o}}^{c,v}$ and $B_{\mathbf{o}}^{c,v}$ of the wave functions is
almost preserved, even though the modulated field does have some effects on
the wave functions.

As the modulated field strength $B_{M}$ is increased to $B_{0}=4$ T, the
frequency, intensity, peak number and selection rules in the optical
absorption spectrum have evident variety, as shown in Fig. 6(b). The
frequencies of $\omega _{\alpha }^{nn^{\prime }}$'s are largely reduced, but
the opposite is true for $\omega _{\beta }^{nn^{\prime }}$'s. The fact that
the $\alpha $ and $\beta $ band-edge states, respectively, approach or
depart from $E_{F}=0$ is the main reason.\ This might lead to an abnormal
relationship of absorption frequencies in that $\omega _{\alpha
}^{nn^{\prime }}$'s with the large $n^{c,v}$'s occur at a lower frequency
than $\omega _{\beta }^{nn^{\prime }}$ with the small $n^{c,v}$'s. For
example, $\omega _{\alpha }^{12}$, $\omega _{\alpha }^{67}$ and $\omega
_{\alpha }^{9\text{ }10}$ are smaller than $\omega _{\beta }^{01}$, $\omega
_{\beta }^{12}$ and $\omega _{\beta }^{23}$, respectively. The ratio between
the intensities of $\omega _{\alpha }^{nn^{\prime }}$ and $\omega _{\beta
}^{nn^{\prime }}$ is further diminished\ with increasing $B_{M}$. This can
be clearly observed in transitions with small $n^{c,v}$'s, i.e., $\omega
_{\beta }^{01}$ is ten times stronger than $\omega _{\alpha }^{01}$ in terms
of the intensity. Moreover, more absorption peaks exist when $B_{M}$ is
increased. In addition to the peaks $\omega _{\alpha }^{nn^{\prime }}$ and $%
\omega _{\beta }^{nn^{\prime }}$ obeying the selection rule $\Delta n=1$,
two extra peaks with $\Delta n=2$ and $3$, $\omega _{\alpha }^{02}$ and $%
\omega _{\alpha }^{03}$, are generated. These two peaks reflect the fact
that the wave functions of the LLs with $n^{c,v}=0$ are drastically changed
by the modulated magnetic field.

\begin{figure}[tbp]
\begin{center}
\includegraphics[]{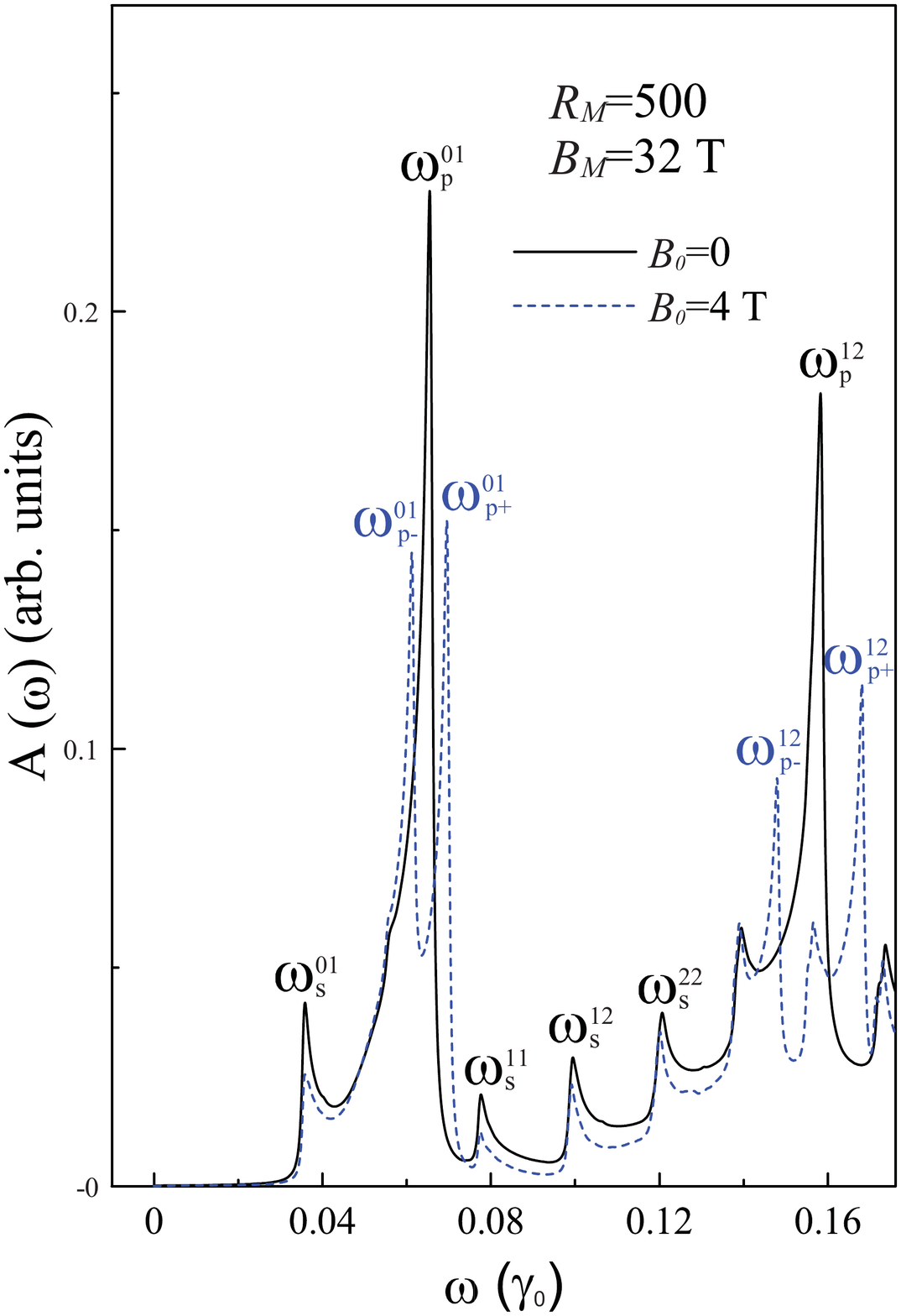}\\[0pt]
\end{center}
\caption{ The low-frequency optical absorption spectra for the $B_{M}>B_{0}$
case corresponding to the composite field $B_{0}=4$ T combined with $%
R_{M}=500$ and $B_{M}=32$ T (dashed blue curve) and the pure modulated
magnetic field $R_{M}=500$ and $B_{M}=32$ T (black curve).}
\label{fig7.eps}
\end{figure}

As the modulated field strength is further raised to $B_{M}=32$ T (red
dashed curves in Fig. 7), the absorption spectrum displays certain important
features in the pure modulated magnetic field where $R_{M}=500$ and $%
B_{M}=32 $ T (black solid curves in Fig. 7). When the absorption peaks are
dominated by the modulated magnetic field, they can be divided into
principal peaks $\omega _{P}^{nn+1}$ and subpeaks $\omega _{s}^{nn^{\prime
}} $. These two kinds of peaks are not only seen in the modulated magnetic
field case, but also appear in the composite magnetic field case as $%
B_{M}\gg B_{0}$. The subpeaks $\omega _{s}^{nn^{\prime }}$ with the
left-hand divergence arise from transitions between $\alpha $\ type
band-edge states (extra band-edge states) with the quantum number $n$ and $%
n^{\prime }$ in the $B_{M}\gg B_{0}$ case (only $B_{M}$ case). The
association with the positions at zero net field strength are almost
identical in both cases. However, we obtain two selection rules, $\Delta n=0$
and $\Delta n=1$, due to the complex overlapping behavior of two subenvelope
functions in the wave function. On the other hand, the principal peaks $%
\omega _{P}^{nn+1}$ with right-hand divergence are much stronger than the
subpeaks. They are attributed to the higher DOS consisting of the $\beta $
type edge-states (specific band-edge states) with the quantum numbers $n$
and $n+1$ in the composite field $B_{M}\gg B_{0}$ (pure modulated field).
These absorption peaks obey the selection rule $\Delta n=1$,\ similarly to
the uniform magnetic field case. The principal peaks, however, possess a
pair of peaks with $\omega _{P-}^{nn+1}$ and $\omega _{P+}^{nn+1}$, which
respectively correspond to two different field strengths, $\left\vert
B_{0}-B_{M}\right\vert =28$ T and $\left\vert B_{0}+B_{M}\right\vert =36$ T,
and thus the difference between the two field strengths leads to distinct
absorption frequencies. For $B_{M}\gg B_{0}$, one can anticipate that the
frequency discrepancy between the pair $\omega _{P-}^{nn+1}$ and $\omega
_{P+}^{nn+1}$ becomes very small and eventually merges into a single peak, $%
\omega _{P}^{nn+1}$, i.e., the absorption spectrum is restored to the status
of the pure modulated magnetic field case.

\begin{figure}[tbp]
\begin{center}
\includegraphics[]{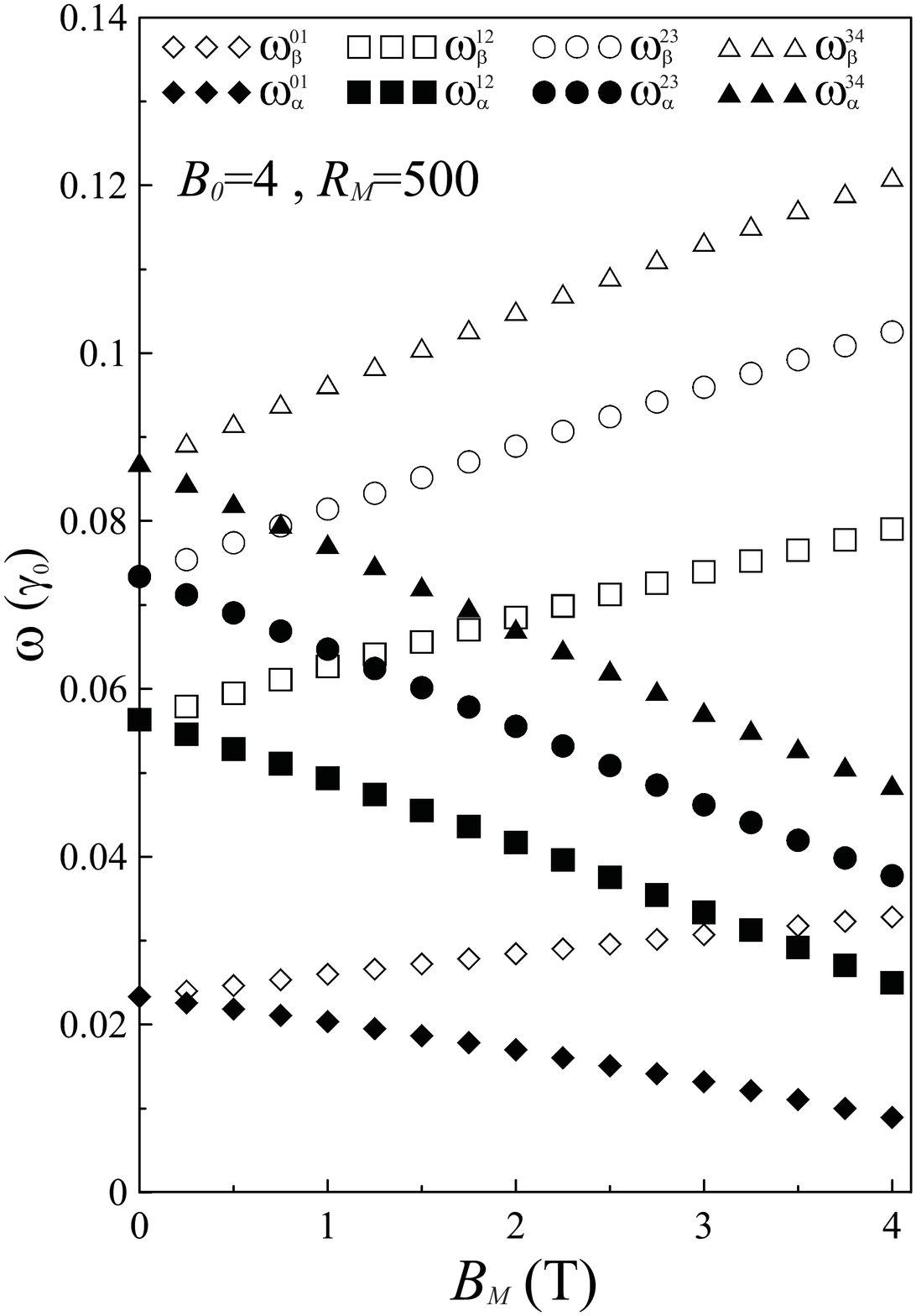}\\[0pt]
\end{center}
\caption{ The dependence of absorption frequencies $\protect\omega _{\protect%
\alpha }^{nn^{\prime }}$ and $\protect\omega _{\protect\beta }^{nn^{\prime
}} $ with $\left\vert \Delta n\right\vert =\left\vert n-n^{\prime
}\right\vert =1$ on the modulated strength $B_{M}$ at $B_{0}=4$ T and $%
R_{M}=500$.}
\label{fig8.eps}
\end{figure}

The absorption frequency dependence on the modulated field strength is shown
in Fig. 8 for $R_{M}=500$ and $B_{0}=4$ T. In the range $B_{M}\leq B_{0}$,
the absorption peaks can be further divided into two categories based on the
different field dependence. The frequencies of $\omega _{\alpha
}^{nn^{\prime }}$ grow with increasing field strength, while the frequencies
of $\omega _{\beta }^{nn^{\prime }}$ decline. Each of the absorption peaks $%
\omega _{\alpha }^{nn^{\prime }}$ and $\omega _{\beta }^{nn^{\prime }}$ has
a linear relationship with $B_{M}$. This reflects the fact that the subband
amplitudes are proportional to $B_{M}$ within the range. It should be noted
that the abnormal relationship of the absorption frequency, the intersection
of $\omega _{\alpha }^{nn^{\prime }}$'s with large $n^{c,v}$'s and $\omega
_{\beta }^{nn^{\prime }}$'s with small $n^{c,v}$'s, occurs at a sufficiently
large modulated field. However, in the higher absorption frequency region or
for the field range $B_{M}>B_{0}$, the linear relationship is broken since
the subband amplitudes are no longer linearly magnified by $B_{M}$.

In addition to the modulated magnetic field, the modulated electric
potential also induces rich features in the magneto-absorption spectra. In
both cases, the four-fold degenerate LLs, except the $n^{c,v}=0$ LL for the
modulated magnetic field case, change into periodic oscillation subbands
with double-degeneracy. The energy dispersions become stronger for an
increased modulated field strength. With regard to the wave functions, the
symmetry breaking of the LL spatial distributions is enhanced as the
modulation potential or the quantum number grows. However, the asymmetry of
the LL wave functions only at $n^{c,v}=0$ is revealed when $B_{M}$ is
comparable to $B_{0}$. This implies that the optical absorption spectrum
simultaneously exhibits the original peaks of $\Delta n=1$ and the extra
peaks of $\Delta n\neq 1$. In the case of the modulated magnetic field, the
extra selection rules, i.e., $\Delta n=2$ and $\Delta n=3$, come into
existence only when $B_{M}$ approximately equals $B_{0}$. On the other hand,
extra selection rules induced by the electric potential would be generated
when the modulation strength is increased. The extra peaks of $\Delta n\neq
1 $ are relatively easily observed at higher frequency.

\section{Conclusion}

The low-frequency optical absorption spectra in the presence of composite
magnetic fields are studied by the generalized tight-binding model and
gradient approximation. By means of controlling the ratio between $B_{M}$
and $B_{0}$, systematic research on the magneto-optical spectra of MG can be
thoroughly carried out.\ Under a uniform magnetic field, many
delta-function-like peaks with uniform intensities exist in the optical
spectrum. Each peak is generated from the LLs and satisfies the specific
selection rule $\Delta n=1$. At $B_{M}\ll B_{0}$, each LL splits into two
periodic oscillatory subbands with two kinds of band-edge states $%
k_{be}^{\alpha }$ and $k_{be}^{\beta }$, except for the LL with $n^{c,v}=0$.
Nevertheless, the spatial symmetry of the wave functions remains unchanged.
The simple relationship between $A_{\mathbf{o}}^{c,v}$ and $B_{\mathbf{o}%
}^{c,v}$ of the wave functions is almost preserved. Optical spectra exhibit
many pairs of square-root-divergent peaks $\omega _{\alpha }^{nn^{\prime }}$
and $\omega _{\beta }^{nn^{\prime }}$ that obey the selection rule $\Delta
n=1$. The two peak frequencies associated with $k_{be}^{\alpha }$ and $%
k_{be}^{\beta }$\ are the same as those generated from the LLs at $%
B_{0}-B_{M}$ and $B_{0}+B_{M}$, respectively. The implication is that a
redshift occurs in the threshold absorption frequency. When $B_{M}$ is
increased to the magnitude of $B_{0}$, the oscillatory subbands display
stronger energy dispersions and greater band curvatures. The strong
oscillatory subbands with different quantum numbers overlap with one
another. The wave functions exhibit broadened and reduced spatial
distributions. In particular, the spatial symmetry of $n^{c,v}=0$ is
severely broken, which thus leads to the inclusion of extra peaks of $\Delta
n\neq 1$ in the absorption spectrum. Moreover, an abnormal relationship of
absorption frequencies exists in the optical spectra, i.e., $\omega _{\alpha
}^{nn^{\prime }}$'s with large $n^{c,v}$'s occur at frequencies lower than $%
\omega _{\beta }^{nn^{\prime }}$ with small $n^{c,v}$'s. As the modulated
field strength is raised to $B_{M}\gg B_{0}$, the optical spectra display
certain features similar to those presented in the case of a pure modulated
field $B_{M}$. Both cases contain the two selection rules $\Delta n=0$ and $%
\Delta n=1$, mainly owing to the complex overlapping behavior of two
subenvelope functions in the wave function. However, in the case $B_{M}\gg
B_{0}$, the principal peaks consist of a pair of peaks that respectively
correspond to the two different field strengths $B_{M}\pm B_{0}$.

Very importantly, the generalized tight-binding model is developed to study
monolayer graphene under various kinds of external fields. These fields can
be uniform magnetic fields, modulated electric fields, modulated magnetic
fields, and composite fields. The Hermitian Hamiltonian matrix used to
determine the magneto-electronic properties becomes very large for the
experimental field strengths. By means of rearranging the tight-binding
functions, it is possible to transform this huge matrix into a band-like one
to improve computational efficiency. The computation time can be further
reduced by utilizing the characteristics of wave function distributions in
the sublattices. In the generalized tight-binding model, the $\pi $%
-electronic structure of MG is exactly solved over the wide energy range $%
\pm 5$ eV, a solution that has proven to be valid even if a magnetic,
electric or composite field is applied. Although the important interlayer
interactions are not used in this case, the developed method is a
generalized treatment more than just perturbations. This model is not only
suitable in theoretical calculations of MG, but can also be extended to
other stacked layer systems, i.e., AA-, AB-, ABC-stacked [53, 55, 57-59]
and bulk systems [60]. The present study should prove very useful for
comprehending other physical properties, such as Coulomb excitations [61-66]
and transport properties [67-71].

\section*{Acknowledgments}

This work is supported by the NSC of Taiwan, under Grant No. NSC
102-2112-M-006-007-MY3.

\end{document}